\documentclass[sigconf]{acmart}

\usepackage{xcolor, colortbl}
\usepackage[inline]{enumitem}

\AtBeginDocument{%
  }
\setcopyright{acmlicensed}
\copyrightyear{2026}
\acmYear{2026}
\acmDOI{XXXXXXX.XXXXXXX}
\acmConference[ICSE JAWs 2026]{Journal Ahead Workshop (JAWs) 2026}{April 12--18,
  2026}{Rio de Janeiro, Brazil}
\acmISBN{978-1-4503-XXXX-X/2018/06}

\begin{document}
\begin{sloppy}

\title{On the Impact of AGENTS.md Files on the Efficiency of AI~Coding~Agents}

\author{Jai Lal Lulla}
\affiliation{%
  \institution{Singapore Management University}
  \city{Singapore}
  \country{Singapore}}
\email{jailal.l.2025@phdcs.smu.edu.sg}

\author{Seyedmoein Mohsenimofidi}
\affiliation{%
  \institution{Heidelberg University}
  \city{Heidelberg}
  \country{Germany}}
\email{s.mohsenimofidi@uni-heidelberg.de}

\author{Matthias Galster}
\affiliation{%
  \institution{University of Bamberg}
  \city{Bamberg}
  \country{Germany}}
\email{mgalster@ieee.org}

\author{Jie M. Zhang}
\affiliation{%
  \institution{King's College London}
  \city{London}
  \country{United Kingdom}}
\email{jie.zhang@kcl.ac.uk}

\author{Sebastian Baltes}
\affiliation{%
  \institution{Heidelberg University}
  \city{Heidelberg}
  \country{Germany}}
\email{sebastian.baltes@uni-heidelberg.de}

\author{Christoph Treude}
\affiliation{%
  \institution{Singapore Management University}
  \city{Singapore}
  \country{Singapore}}
\email{ctreude@smu.edu.sg}

\renewcommand{\shortauthors}{Lulla et al.}

\begin{abstract}
AI coding agents such as Codex and Claude Code are increasingly used to autonomously contribute to software repositories.
However, little is known about how repository-level configuration artifacts affect operational efficiency of the agents.
In this paper, we study the impact of \texttt{AGENTS.md} files on the runtime and token consumption of AI coding agents operating on GitHub pull requests.
We analyze 10 repositories and 124 pull requests, executing agents under two conditions: with and without an \texttt{AGENTS.md} file. We measure wall-clock execution time and token usage during agent execution. Our results show that the presence of \texttt{AGENTS.md} is associated with a lower median runtime ($\Delta 28.64$\%) and reduced output token consumption ($\Delta 16.58$\%), while maintaining a comparable task completion behavior.
Based on these results, we discuss immediate implications for the configuration and deployment of AI coding agents in practice, and outline a broader research agenda on the role of repository-level instructions in shaping the behavior, efficiency, and integration of AI coding agents in software development workflows.
\end{abstract}

\ccsdesc[500]{Software and its engineering~Software configuration management and version control systems}

\keywords{AI coding agents, AGENTS.md, pull requests, efficiency}

\received{23 January 2026}

\maketitle

\section{Introduction}

AI-assisted software development has evolved rapidly from tools that support individual programming actions to systems that can autonomously carry out multi-step development tasks. Large language models are now routinely used for code generation, testing, repair, review, and documentation, covering substantial portions of the software development lifecycle~\cite{fan2023large, hou2024large, jiang2024survey}. Building on these capabilities, recent AI coding agents, such as OpenAI Codex and Claude Code, can navigate repositories, reason over multiple files, execute commands, and submit pull requests with limited human intervention, effectively acting as autonomous contributors rather than passive assistants~\cite{yang2024swe, roychoudhury2025agentic}.

As these agents operate with increasing autonomy, their behavior depends not only on model capabilities but also on the contextual information provided by the repository. Previous research has largely focused on interaction-level prompt engineering and agent planning strategies~\cite{sahoo2024systematic}, while less attention has been paid to persistent, repository-level artifacts that encode project-specific knowledge. In practice, developers have begun to introduce agent context files such as \texttt{AGENTS.md} or \texttt{CLAUDE.md} that serve as ``READMEs for agents,'' specifying architecture, build commands, coding conventions, and operational constraints~\cite{mohsenimofidi2026context}. The \texttt{AGENTS.md} format, for example, has been adopted by more than 60,000 repositories to date~\cite{agentsAGENTSmd}. These files shift agent guidance from ephemeral prompts toward version-controlled, inspectable, and collaboratively maintained configuration artifacts.

Recent empirical studies have established agent context files as a widespread and evolving phenomenon in open-source software. \citeauthor{chatlatanagulchai2025agent} show that such files are actively maintained, structurally consistent, and heavily focused on functional instructions such as build, testing, and implementation details, while non-functional concerns such as performance and security are comparatively rare~\cite{chatlatanagulchai2025agent}. Similarly, \citeauthor{mohsenimofidi2026context} document the emergence of agent context files as a mechanism for context engineering in agentic coding workflows, highlighting their role in shaping agent behavior across repositories~\cite{mohsenimofidi2026context}. Together, these studies establish agent context files as a new class of software artifacts that play a central role in agentic development.

Although prior work characterizes the structure, content, and evolution of agent context files, their concrete impact on the behavior of AI coding agents remains largely unexplored. In particular, there is little empirical evidence on how repository-level instructions influence the behavior of AI coding agents once they are deployed in real development settings. As agents become more deeply integrated into continuous development workflows, understanding their behavioral and operational implications becomes increasingly important, with potential consequences for cost, scalability, and workflow integration.

In this paper, we take a first step toward understanding the impact of \texttt{AGENTS.md} files on AI coding agents by conducting an empirical study on real GitHub pull requests. Inspired by recent work on cost-efficient software engineering agents~\cite{guo2026eet} and sustainable software engineering~\cite{venters2023sustainable}, we compare agent executions with and without an \texttt{AGENTS.md} file, focusing on the operational efficiency of agents as a concrete and measurable aspect of agent behavior. Specifically, we analyze wall-clock execution time and token consumption, which directly relate to agent latency and computational cost in practice.
Beyond this focused investigation, we outline a broader research roadmap that situates agent efficiency as one dimension within a more comprehensive study of how \texttt{AGENTS.md} files influence the behavior and integration of AI coding agents in software development workflows.

\section{Background and Related Work}

Repository-maintained instruction files are increasingly being adopted as a mechanism to align AI coding agents with project-specific norms and expectations. In particular, the use of a repository-level \texttt{AGENTS.md} file has emerged in multiple agent tooling ecosystems. OpenAI Codex documents layered instruction discovery that incorporates repository-level \texttt{AGENTS.md} files \cite{openai_agentsmd}, while GitLab Duo similarly describes project-level \texttt{AGENTS.md} files to scope and guide agent behavior \cite{gitlab_agentsmd}. Together, these efforts suggest that persistent, developer-curated instruction artifacts are becoming a practical and shared interface for shaping agent behavior in real software repositories.

Despite this emerging adoption, we are not aware of empirical studies that isolate the impact of adding an \texttt{AGENTS.md} file on AI coding agent behavior or efficiency. In particular, no prior work has evaluated how such instruction files affect token usage or wall-clock completion time under a paired, same-task/same-repository evaluation design. This gap motivates the present study.

Previous work has examined the performance and efficiency of LLM-based autonomous coding agents in repository-level tasks. Benchmarks such as \emph{SWE-bench} operationalize this setting using real GitHub issues paired with executable repository snapshots, enabling controlled comparisons of autonomous agents based on issue resolution success and test pass rates \cite{jimenez2024swe-bench}. Building on this benchmark, recent agentic systems demonstrate that agent–repository and agent–tool interface design can significantly affect both performance and resource usage. For example, SWE-agent emphasizes agent–computer interface design and context management for repository navigation, editing, and command execution~\cite{yang2024swe}. \emph{AutoCodeRover} similarly combines LLM reasoning with structured search and debugging signals to reduce cost and improve effectiveness on SWE-bench variants~\cite{autocoderover2024}. These works motivate measuring not only task success, but also practical deployment costs such as wall-clock completion time and token usage.

Recent work has also begun to characterize developer-provided, persistent context artifacts for AI coding agents, including files such as \texttt{copilot-instructions.md} and \texttt{CLAUDE.md} \cite{chatlatanagulchai2025agent, jiangnam2025context, mohsenimofidi2026context}. However, we are not aware of studies that \emph{isolate} the effect of introducing an \texttt{AGENTS.md} file on agent efficiency, holding tasks, repositories, and agent architecture constant.

\section{Study Design}

This study investigates how the presence of an \texttt{AGENTS.md} file affects the performance of autonomous AI coding agents when executing development tasks. We model a standard developer workflow: an agent receives a task description akin to a GitHub issue (``ticket''), performs repository-local actions (e.g., bug fixes, feature additions, refactors), and produces a code change intended to match an expected outcome.


We focus on the \emph{efficiency} of AI coding agents, defined as the computational resources required to complete a development task. This leads to the following research question:


\begin{description}[style=multiline, leftmargin=8mm]
    \item[RQ:] Does the presence of an \texttt{AGENTS.md} file in the repository root reduce the resources required by an autonomous AI coding agent to complete a development task?
\end{description}
    
We operationalize ``resources'' using (i) token usage and (ii) wall-clock time-to-completion measured during each task run. We compare these metrics between paired runs performed with and without an \texttt{AGENTS.md} file in the repository root, keeping the task and repository snapshot constant.



\subsection{Data Collection and Analysis}

\subsubsection{Agent Selection}

The agent used in this study is \emph{OpenAI Codex}. At the time of experimentation, the latest available Codex model was \texttt{gpt-5.2-codex}~\cite{openaiCodexModels}, which we use consistently across all experiments. We selected Codex because it is specifically designed for software engineering tasks, supports repository-scale context and tool use, and is representative of production-grade AI coding agents used in practice.
Moreover, \texttt{AGENTS.md} was first used with Codex before becoming an open format.
Evaluating the effect of \texttt{AGENTS.md} across different agent systems and model families is an important direction for future work and part of our ongoing research agenda (Section~\ref{sec:research_roadmap}).

We implemented a lightweight Python wrapper to interface with the Codex CLI and to automate task setup and metric collection, available in our online appendix~\cite{lulla2026agentsmdappendix}.
In the following, the term \emph{agent} refers to this Codex-based agent configuration.

\subsubsection{Repository Sampling and Inclusion Criteria}

We begin from a corpus of repositories sampled in previous work \cite{mohsenimofidi2026context} that analyzed the adoption of agent instruction files, such as \texttt{AGENTS.md}. In that corpus, repositories may contain (i) multiple instruction files of different names, (ii) multiple copies in different subdirectories, or (iii) only one file at the root. In this study, we focus on the simplest configuration: repositories that contain one \texttt{AGENTS.md} file only at the repository root. This configuration minimizes confounding effects from overlapping or conflicting instruction files and enables clearer attribution of observed agent behavior to a single, well-defined source of repository-level guidance. Applying this constraint yields 89 repositories (from 132 total).

To ensure that our evaluation focuses on \texttt{AGENTS.md} files that plausibly provide actionable project context to an AI coding agent, we further restrict the dataset based on the content of the instruction files. Concretely, we filter the root \texttt{AGENTS.md} files using content categories derived from the taxonomy developed in~\cite{mohsenimofidi2026context}. This taxonomy characterizes \texttt{AGENTS.md} content along dimensions such as coding conventions and best practices, architecture and project structure, project description, testing instructions, and security. We retain only those files that contain information related to (i) conventions and best practices, (ii) architecture and project structure, and (iii) project description. These categories were selected because they are the most common ones~\cite{mohsenimofidi2026context} and they capture core project knowledge that developers typically need in order to understand and contribute to a codebase~\cite{he2025llmasajudgesoftwareengineeringliterature}.

The classification of \texttt{AGENTS.md} files was performed using an LLM (\texttt{gpt-oss-120b}) according to the above criteria. The model was run using Ollama~\cite{ollamaLib}, followed by manual verification of the filtered results. After applying this filtering step, we retain 26 repositories, each containing a qualifying root \texttt{AGENTS.md} file.

\subsubsection{Pull Request Selection and Task Construction}

From the 26 repositories, we randomly sample 10 and select up to 15 merged pull requests (PRs) from each. The cap reflects resource constraints while still enabling coverage across repositories. To obtain task instances that can be executed repeatedly and comparable across repositories, we restrict PRs to small-scope, code-changing contributions and ensure that each PR post-dates the introduction of \texttt{AGENTS.md} in the repository. Each PR satisfies the following criteria:
\begin{enumerate*}[topsep=0pt]
  \item Size constraint: total additions + deletions $\leq$ 100 LoC;
  \item Scope constraint: $\leq$ 5 modified files;
  \item Status: merged PRs only;
  \item Temporal constraint: PR created and merged after the introduction of \texttt{AGENTS.md} in that repository;
  \item Change type constraint: PR modifies code files only (excluding documentation and configuration changes).
\end{enumerate*}

The size and scope constraints reduce the variance from large refactorings and keep agent runs tractable and repeatable. The change-type constraint avoids skew from PRs that do not reflect code changes (e.g., documentation-only updates or version bumps). Together, these restrictions help isolate the effect of \texttt{AGENTS.md} in this initial study. Relaxing them to include larger changes and a broader range of PR types is part of our longer-term research agenda (Section~\ref{sec:research_roadmap}).


\subsubsection{Reconstructing Pre-PR Repository State}

To recreate realistic development conditions, we reconstruct each repository to the state immediately before the selected PR was merged. The agent is then tasked with recreating the PR’s changes from this pre-merge state. Concretely, for each PR we:

\begin{enumerate}[topsep=0pt]
    \item Check out the repository at the pre-merge commit.
    \item Extract the \texttt{AGENTS.md} version that existed at that commit.
    \item Run the agent on the pre-merge repository snapshot.
\end{enumerate}

This produces an evaluation setup grounded in an actual, historically merged change, where the repository contents and instruction file (when present) exactly match what a developer or agent would have observed at that point in time.

\subsubsection{Issue Generation for PRs Lacking Usable Descriptions}

Many PRs do not contain sufficient natural-language context (e.g., empty body, ``fix bug'' titles, missing linked issues). To provide consistent and informative task input, we generate a GitHub-issue-style task statement for each PR using a local LLM (\texttt{gpt-oss-120b}). The model is prompted with:
\begin{enumerate*}[topsep=0pt]
    \item the PR diff (patch) and
    \item the repository structure at pre-merge state (e.g., file tree).
\end{enumerate*}

The output is a structured task description that resembles a GitHub issue (problem statement, expected behavior, constraints, and acceptance criteria). This step standardizes the agent's input format across PRs and reduces variance introduced by incomplete PR metadata~\cite{chaparro2019issues}. The generated task descriptions are available online~\cite{lulla2026agentsmdappendix}.

\subsubsection{Experimental Conditions}

We run the agent on each task instance under two conditions: 

\begin{itemize}[topsep=0pt]
    \item \textbf{With} \texttt{AGENTS.md}: The repository snapshot includes the extracted root \texttt{AGENTS.md} from the pre-merge commit.
    \item \textbf{Without} \texttt{AGENTS.md}: The same snapshot is used, but the \texttt{AGENTS.md} file is removed (all other files unchanged).
\end{itemize}

In both conditions, the agent is provided with the same task input, namely the GitHub-like issue generated earlier for each pull request. The agent then produces code changes corresponding to the original PR. The resulting setup constitutes a \textbf{paired within-task design} that controls for repository, task, and codebase state, while varying only the presence of \texttt{AGENTS.md}.

\subsubsection{Running the Experiments}

To minimize confounding factors and ensure repeatability, experiments were conducted in isolated Docker environments at the repository level. For each repository, we instantiated a fresh container and cloned the target repository into a temporary working directory. Each task instance was executed by checking out the corresponding pre-merge commit within the container. After task completion, the repository was reset to a clean state (discarding all local changes), and the workspace was cleaned before checking out the commit associated with the next task. The code used to run the experiments, including Dockerfiles, is provided as part of the supplementary material~\cite{lulla2026agentsmdappendix}.

The agent was granted access only to this sandboxed environment and could modify files exclusively within the container. No state (e.g., caches, artifacts, or intermediate files) was reused across tasks beyond the version-controlled repository contents, ensuring that every task began from an identical repository snapshot.


\subsubsection{Metrics}

We collected the following operational performance metrics during each run:

\begin{itemize}[topsep=0pt]
    \item \textbf{Token usage}: Total tokens consumed, comprising input tokens, cached input tokens, and output tokens.
    \item \textbf{Time-to-completion}: Wall-clock time for the agent to produce its final output (in seconds).
\end{itemize}

A comprehensive evaluation of the output quality, e.g., the semantic correctness or the functional equivalence to the merged PR, is beyond the scope of this paper. However, it is part of our research roadmap described in Section~\ref{sec:research_roadmap}. Nevertheless, to ensure that the observed efficiency differences are not simply due to agents producing degenerate or trivially incomplete output, we performed a manual sanity check. Specifically, we randomly sampled \textbf{50} PR tasks and inspected the corresponding agent outputs, comparing them against the human-written merged pull requests, to confirm that they resulted in non-empty, non-trivial code changes consistent with the intended task, rather than aborted runs or random edits. Although this sanity check does not constitute a full correctness evaluation, it provides confidence that the efficiency measurements reported in this paper are not driven by obvious failures or reductions in task execution.


\section{Results}

\begin{table}[t]
\centering
\small
\caption{Resource Usage With and Without \texttt{AGENTS.md}}
\label{tab:rq_1-metrics_data}
\begin{tabular}{lrrrr}
\hline
\textbf{Metric} & \textbf{Without} & \textbf{With} & \textbf{Diff} & \textbf{$\Delta$\%}
 \\
\hline
\multicolumn{5}{l}{\cellcolor{gray!10}\textbf{Wall-Clock Time (s)$^{*}$}} \\
\hline
\quad Mean & 162.94 & 129.91 & 33.03 & 20.27\% \\
\quad Median & 98.57 & 70.34 & 28.23 & 28.64\% \\
\quad Std Dev & 182.24 & 136.84 & 45.40 & 24.91\% \\
\hline
\multicolumn{5}{l}{\cellcolor{gray!10}\textbf{Input Tokens}} \\
\hline
\quad Mean & 353{,}010.01 & 318{,}651.51 & 34{,}358.50 & 9.73\% \\
\quad Median &  116{,}609.00 & 120{,}587.00 & -3{,}978.00 & -3.41\% \\
\quad Std Dev & 654{,}603.95 & 510{,}776.51 & 143{,}827.43 & 21.97\% \\
\hline
\multicolumn{5}{l}{\cellcolor{gray!10}\textbf{Cached Input Tokens}} \\
\hline
\quad Mean & 328{,}877.31 & 296{,}078.73 & 32{,}798.58 & 9.97\% \\
\quad Median & 103{,}424.00 & 104{,}448.00 & -1{,}024.00 & -0.99\% \\
\quad Std Dev & 632{,}622.27 & 494{,}157.89 & 138{,}464.38 & 21.89\% \\
\hline
\multicolumn{5}{l}{\cellcolor{gray!10}\textbf{Output Tokens}$^{*}$} \\
\hline
\quad Mean & 5{,}744.81 & 4{,}591.46 & 1{,}153.35 & 20.08\% \\
\quad Median & 2{,}925.00 & 2{,}440.00 & 485.00 & 16.58\% \\
\quad Std Dev & 6{,}987.74 & 5{,}161.67 & 1{,}826.06 & 26.13\% \\
\hline
\multicolumn{5}{l}{\cellcolor{gray!10}\textbf{Total Tokens}} \\
\hline
\quad Mean & 687{,}632.13 & 619{,}321.70 & 68{,}310.43 & 9.93\% \\
\quad Median & 223{,}707.00 & 226{,}582.00 & -2{,}875.00 & -1.29\% \\
\quad Std Dev & 1{,}293{,}176.16 & 1{,}009{,}338.80 & 283{,}837.36 & 21.95\% \\
\hline
\end{tabular}
{\footnotesize
\par\noindent
$^{*}$ Statistically significant difference, Wilcoxon signed-rank test ($p < 0.05$).
}
\vspace{-4mm}
\end{table}


\paragraph{Token usage} Providing an \texttt{AGENTS.md} file reduces generation cost. Mean output token usage decreases from 5{,}744.81 tokens (without \texttt{AGENTS.md}) to 4{,}591.46 tokens (with \texttt{AGENTS.md}), a reduction of 1{,}153.35 tokens ($\approx 20.08\%$). Median output tokens decrease more modestly, from 2{,}925.00 to 2{,}440.00 (485 tokens, $\approx 16.58\%$). Mean input and cached input tokens also decline slightly (353{,}010.01 to 318{,}651.51; $9.73\%$ and 328{,}877.31 to 296{,}078.73; $9.97\%$, respectively), with medians essentially unchanged or slightly higher. The larger reduction in mean output tokens compared to the median suggests that \texttt{AGENTS.md} primarily reduces token usage in a small number of very high-cost runs, rather than uniformly lowering token consumption across all task instances.

\paragraph{Wall-clock time-to-completion} Agents provided with an \texttt{AGENTS.md} file complete tasks faster than agents operating without it. Mean wall-clock time-to-completion decreases from 162.94s (without \texttt{AGENTS.md}, std dev 182.24s) to 129.91s (with \texttt{AGENTS.md}, std dev 136.84s), an absolute reduction of 33.03s ($\approx 20.27\%$). Median completion time shows a similar reduction, decreasing from 98.57s to 70.34s (28.23s, $\approx 28.64\%$). The close alignment between mean and median improvements indicates that the reduction is not driven solely by a small number of extreme runs, but reflects a general shift toward faster task completion.


\begin{figure}
    \includegraphics[width=0.85\linewidth]{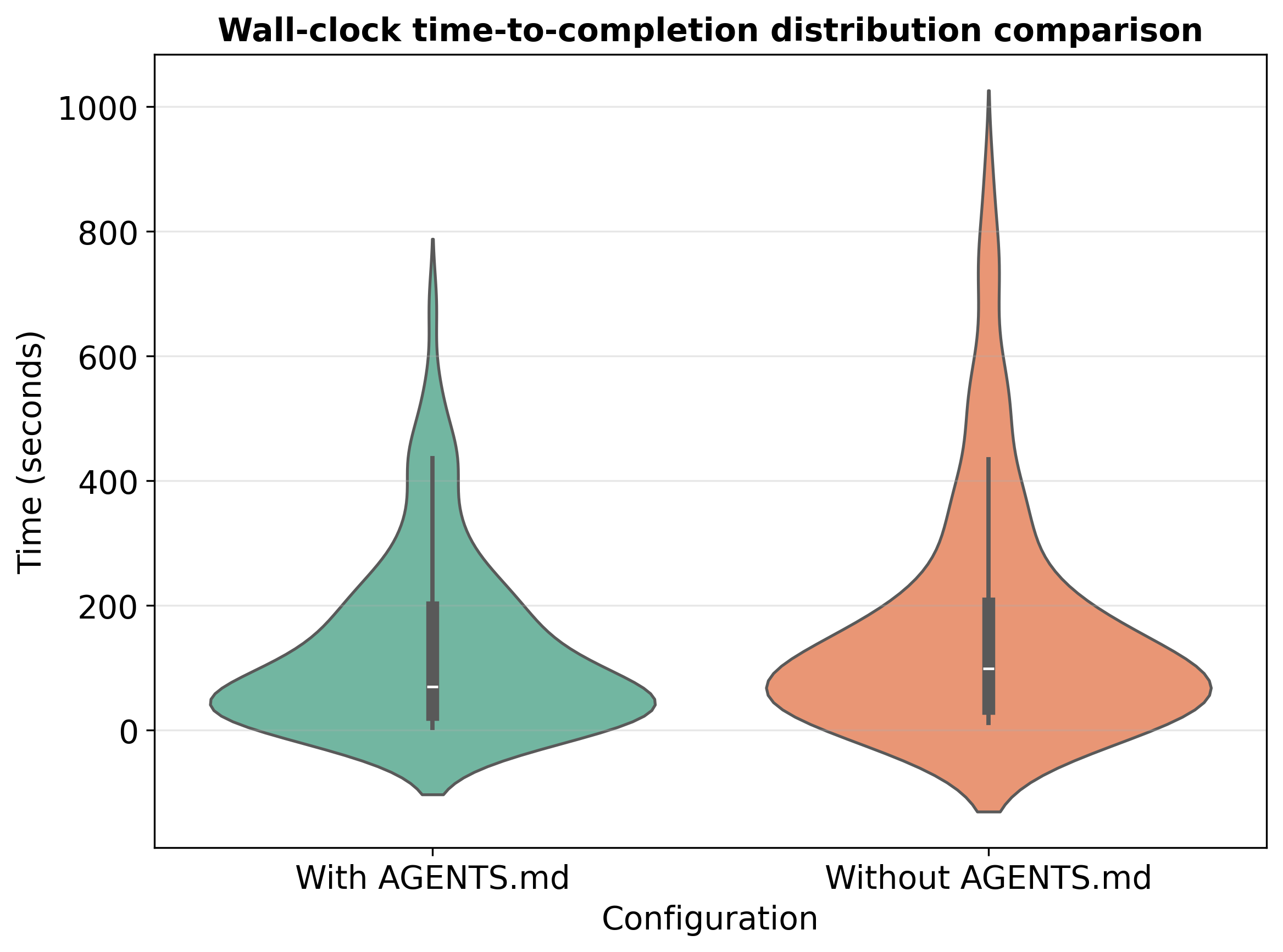}
    \caption{Wall-clock time-to-completion distributions for agent runs with and without \texttt{AGENTS.md}}\label{fig:time_plot}
    \vspace{-6mm}
\end{figure}

\section{Research Roadmap}\label{sec:research_roadmap}

This study represents an initial step towards understanding how \texttt{AGENTS.md} files influence the behavior of autonomous AI coding agents. Our results provide early evidence that such files can affect agent efficiency in realistic development tasks. At the same time, the study focuses on a sampled subset of repositories, pull requests, and agent configurations, which naturally motivates the next steps.

A first direction for future work is to expand the empirical scope of the study. Although we control for task, repository state, and agent configuration through a paired design, observed effects may still depend on agent stochasticity, the specific agent framework and model used, and the characteristics of the selected tasks. Replicating the study across additional repositories, larger and more diverse pull requests, and multiple agent systems and model families will help assess the robustness and generality of the observed efficiency effects. Similarly, relaxing current constraints on task size and scope will allow us to examine whether the influence of \texttt{AGENTS.md} extends to larger refactorings, multi-module changes, and more complex development scenarios.

A second direction is to investigate dimensions of agent behavior beyond efficiency. In this study, we focus on token usage and wall-clock time as measurable indicators of cost. However, these metrics do not capture whether agent-produced changes are correct, maintainable, or aligned with developer intent. Future work will therefore incorporate correctness and alignment evaluations, for example, by comparing agent-generated changes against the original merged pull requests using automated checks and structural similarity analyzes. Beyond a binary treatment of \texttt{AGENTS.md} as present or absent, we also plan to examine how properties of these files, such as specificity, organization, and the inclusion of workflow guidance, relate to agent outcomes. For example, we speculate that some of the efficiency gains reported in this paper arise because \texttt{AGENTS.md} files describe repository structure and conventions upfront, reducing the need for agents to infer project organization through exploratory navigation. Analyzing agent execution traces will help explain why \texttt{AGENTS.md} files lead to more efficient generation, for example, by fewer planning iterations, reduced exploratory navigation, and fewer repeated requests to the underlying model.

In summary, our results provide initial empirical evidence that repository-level instruction files can have measurable operational effects on autonomous AI coding agents. In particular, we show that the presence of a root \texttt{AGENTS.md} file is associated with reduced token usage and faster task completion on real pull requests. These findings position \texttt{AGENTS.md} as a practical repository-level mechanism for shaping agent behavior and motivate further investigation of its role in agent efficiency, alignment, and integration within software development workflows.

\bibliographystyle{ACM-Reference-Format}
\bibliography{agents.md-impact-jaws}

\end{sloppy}
\end{document}